\documentclass[epj]{svjour}
\usepackage{graphics}
\begin{document}
\title{Interpreting Near-side Correlations }
\subtitle{and the RHIC Beam Energy Scan}
\author{Paul Sorensen
}                     
\institute{Brookhaven National Laboratory, Upton, New York}
\date{Received: date / Revised version: date}
%
\abstract{ Recent data from heavy ion collisions at RHIC show strong
  near-side correlations extending over several units of
  rapidity. This ridge-like correlation exhibits an abrupt onset with
  collision centrality. In this talk, I argue that the centrality and
  beam-energy dependence of these near-angle correlations could
  provide access to information about the Quark Gluon Plasma phase
  boundary and the Equation of State of nuclear matter. A
  beam-energy-scan at RHIC will better reveal the true source of these
  correlations and should be a high priority at RHIC.
} 
\maketitle
\section{Introduction}
\label{intro}
Correlations and fluctuations have long been considered a promising
signature for Quark Gluon Plasma (QGP) formation in heavy-ion
collisions~\cite{fluctuations}. Early proposals for QGP searches
suggested searching for a non-monotonic dependence of fluctuations on
variables that can be related to the energy density created in the
system --- e.g.  center-of-mass energy or collision centrality --- the
expectation being that above some energy density threshold, a phase
transition to QGP would occur. The presence of the phase transition to
QGP would then lead to a change in fluctuations and correlations.

These early expectations regarding finite temperature QCD are bourne
out by lattice QCD calculations~\cite{lattice1}. The calculations show
that for temperatures above a critical value of 195 MeV, a QGP is
formed. Lattice calculations also show that baryon number,
strangeness, and charge fluctuations are all enhanced near the
critical temperature $T_C$~\cite{lattice2}. As such, correlations and
fluctuations remain a topic of interest in heavy-ion collisions.

Data from the experiments at RHIC indeed reveal interesting features
in the two-particle correlation
landscape \cite{onset,ridgedata}. Specifically, it has been found that
correlation structures exist that are unique to Nucleus-Nucleus
collisions. While two-particle correlations in p+p and d+Au collisions
show a peak narrow in azimuth and rapidity, the near-side peak in
Au+Au collisions broadens substantially in the longitudinal direction
and narrows in azimuth. The longitudinal width of the correlation
appears to depend on the $p_T$ of the particles. An analysis of the
width of the peak for particles of all $p_T$ finds the correlation
extends across nearly 2 units of pseudo-rapidity $\eta$. When
triggering on higher momentum particles ($p_T>2$~GeV/c for example),
the correlation extends beyond the acceptance of the STAR detector
($\Delta\eta<2$) and perhaps as far as $\Delta\eta=4$ as indicated by
preliminary PHOBOS data. Furthermore, STAR has found that these
correlations show an abrupt onset as a function of
centrality~\cite{onset}. Comparing measurements at 200 and 62.4 GeV,
STAR has shown that the onset happens at the same value of transverse
particle density for the different energies, suggesting the onset of
the long range correlations may be related to a critical energy
density.

\section{The Ridge}
\label{sec:1}

\begin{figure*}[htb]
\resizebox{0.5\textwidth}{!}{\includegraphics{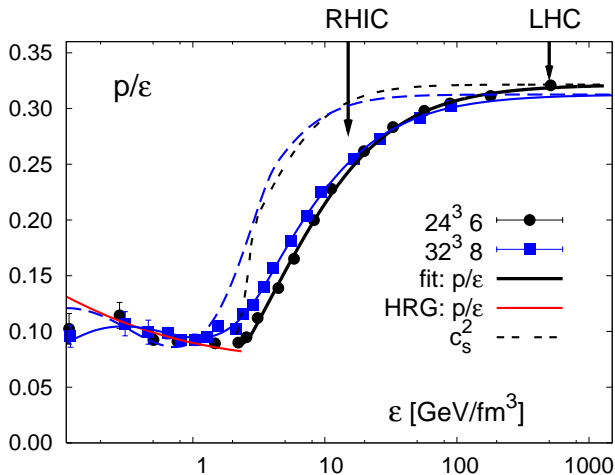}}
\hspace{0.1\textwidth}
\resizebox{0.4\textwidth}{!}{\includegraphics{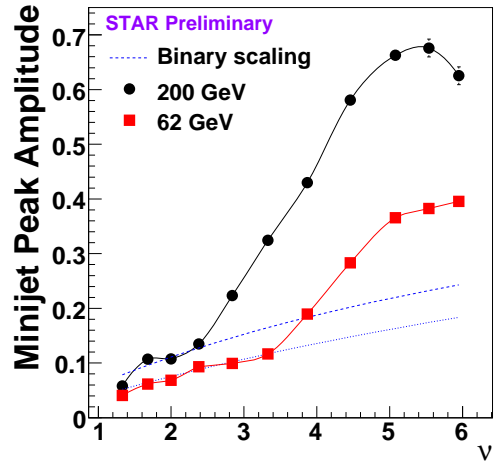}        }
\caption[]{ Left panel: Lattice QCD calculations of the QCD equation
  of state. Right panel: A preliminary STAR figure showing the
  amplitude of the near-side ridge-like peak as presented in
  Ref.~\cite{DGMV} by special permission from the STAR
  collaboration~\cite{onset}.}
\label{fig1}
\end{figure*}

The ridge is a long-range correlation unique to A+A collisions that
exhibits an abrupt onset with increasing collision centrality. Phrased
this way, we could conclude that this is the long saught after
``smoking gun'' of QGP formation. But the excitement one would expect
from such a discovery has been tempered due to conflicting
interpretations of the nature of these correlations. Disagreement
exists as to whether the correlations are related to QGP formation or
whether they in fact disprove the existance of a thermalized
medium. Questions surrounding the ridge-like correlations include: are
the correlations related to non-perturbative multi quark or gluon
effects on minijets in Au+Au collisions?~\cite{minij} are they related
to soft gluons radiated by hard partons traversing the overlap
region?~\cite{Maj} are they related to initial spatial correlations in
the system converted to momentum-space correlations by a radial hubble
expansion?~\cite{radflow} to beam-jets also boosted by the radial
expansion?~\cite{beamjets} or to viscous effects?~\cite{visc} and do
these correlations disprove the assumption of a system thermalized
over some extended region?~\cite{onset} These questions still remain
to be answered.

\begin{figure}
\hspace{0cm}
\resizebox{0.5\textwidth}{!}{\includegraphics{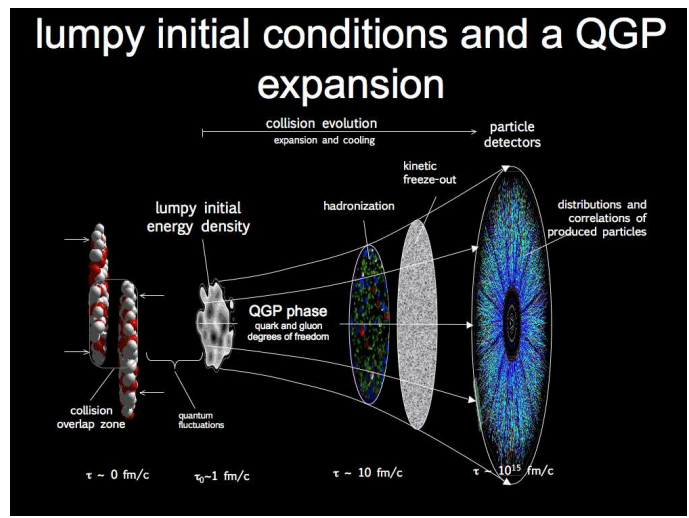}}
\caption{A schematic diagram of the evolution of a heavy-ion
  collision. Possible correlations and fluctuations in the initial
  conditions are emphasized.}
\label{fig2}       
\end{figure}

If the ridge is related to the translation of spatial correlations
into momentum space correlations through radial flow, then the onset
of the ridge could be related to a sharp rise in the pressure over
energy-density at the critical energy density for QGP formation. At
that energy density, the liberation of color degrees of freedom in the
system could lead to an increase in the pressure which in turn could
lead to the flow that makes it possible to image the underlying
spatial correlations in momentum space. It's not clear that
hydrodynamic models will be able to reproduce such effects and the
process by which the QGP transforms initial spatial correlations into
momentum space correlations could be quite different than envisioned
in such models. Fig.~\ref{fig1} shows lattice QCD calculations of the
equation of state~\cite{lattice1} on the left and the ridge amplitude
at 200 and 62.4 GeV vs centrality measure $\nu=2N_{binary}/N_{part}$
on the right. When plotted vs transverse particle density related to
Bjorken energy density, STAR finds that the abrupt increase in the
ridge happens at the same density for both 200 and 62.4 GeV
collisions~\cite{onset}; not the same $\nu$, $N_{part}$, or $N_{bin}$
but the same $\frac{1}{S}\frac{dN}{dy}$. Fig.~\ref{fig2} shows a
schematic illustration of the expansion after a heavy-ion collision
with an emphasis on the lumpiness of the initial conditions.

Above the transition, the ridge amplitude grows faster than
$N_{binary}$ scaling. One analysis finds that when measureing the
correlations between the leading and subleading hadrons in an event,
the area of the ridge scales with the area of the
background~\cite{nav}; perhaps suggesting that the ridge correlation
is related to a bulk correlation (a global correlation between all
particles). It is also found that the baryon to meson ratio in the
ridge is similar to the bulk and that the $p_T$ spectrum of the ridge
is softer than for the jet-cone correlation. In fact, the only feature
of the ridge that matches that of the jet cone, is that it is centered
at $\Delta\phi=0$. For this reason, one may reasonably doubt whether
the ridge is related to hard-scattering.

There are also observations that may prefer an explanation related to
mini-jets or hard scattering. The ridge persists even when
correlations are formed with trigger particles well into the
fragmentation region. Also, the yield of the away-side ridge follows
that of the near-side very closely; perhaps indicating
back-to-back dijet-like correlations~\cite{onset}.

\section{Energy Dependence}
\label{sec:2}

A more extensive beam energy scan can help determine if and how the
abrupt onset of the ridge is related to the onset of deconfinment. A
beam-energy-scan has been proposed at RHIC and technical preparations
have been made to collide beams at $\sqrt{s_{_{NN}}}$ as low as 5
GeV~\cite{escan}. In a recent test run, the STAR collaboration was
able to gather ~3000 good events at $\sqrt{s_{_{NN}}}=9.2$ GeV after
just several hours of beam-time. Collider experts anticipate
increasing these event-rates by approximately a factor of 5. The
number of events required to carry out the ridge analysis shown in
Fig.~\ref{fig1} (right) is on the order of 10 million events. These
data samples can be achieved in less than a week of running for
energies above $\sqrt{s_{_{NN}}}$=25 GeV. At 12.3 GeV, 10 million
events will require approximately four weeks of running to
acquire. This makes an energy scan of ridge phenomenology from 12.3
GeV to 62.4 GeV feasible at RHIC.

Ref.~\cite{DGMV} proposes an explanation for the ridge based on Glasma
flux-tubes. The flux-tubes themselves would not yield a narrow
azimuthal correlation but if they are radially boosted, the emitted
particles can be collimated in azimuth leading to a ridge like
structure. This explanation of the ridge yields a prediction for the
energy dependence of the ridge amplitude. For example,
$\frac{\Delta\rho}{\sqrt{\rho_{ref}}} \propto
\frac{1}{\alpha_{s}(Q_{s})}$ so that the amplitude of the ridge should
be governed by the centrality and energy dependence of $Q_{s}$
modulated by the effectiveness of the space-momentum correlation. In
Ref.~\cite{DGMV}, blast-wave model fits were used to determine the
mean flow velocity. Since the blast-wave fit parameters vary smoothly
with centrality, it is not possible for that implementation of the
flux-tube ridge model to reproduce the abrupt onset of the ridge. The
blast-wave fits to the final hadron spectra do not necessarily
accurately reflect the dynamics of the collision evolution however, so
the lack of an abrupt transition in the model may simply reflect a
weakness of the blast-wave parameterization and of the assumption that
the QGP evolution is well described by hydrodynamics.

\begin{figure}
\resizebox{0.49\textwidth}{!}{\includegraphics{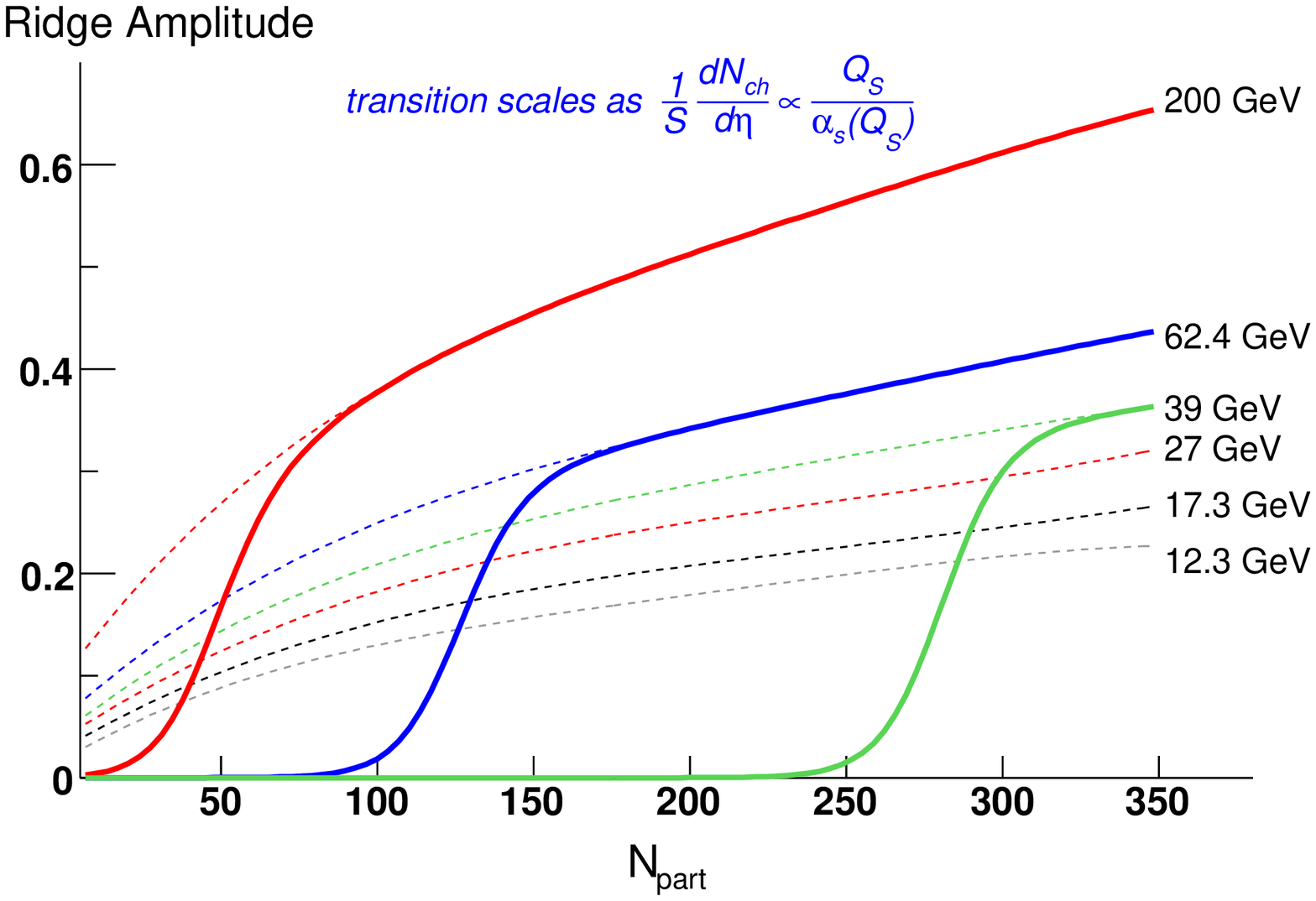}}
\resizebox{0.49\textwidth}{!}{\includegraphics{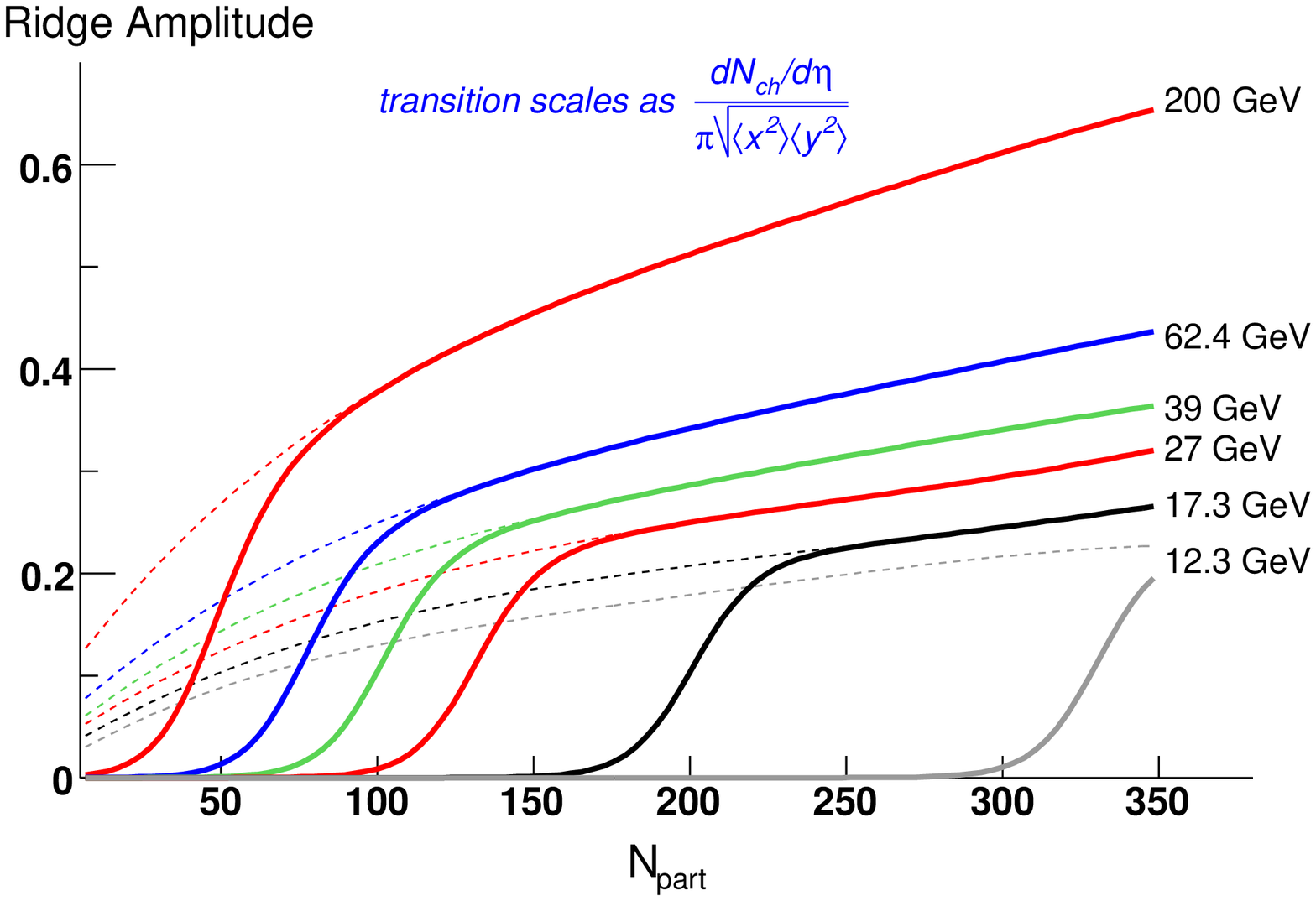}}

\caption{The ridge amplitude from a model of glasma flux-tubes and
  radial flow. The dashed lines are based on flux-tubes and a
  blast-wave model. The solid curves in the top panel include ad-hoc
  abrupt transitions at fixed values of $\frac{1}{S}\frac{dN}{dy}$ as
  suggested by preliminary STAR measurements. The exact location of
  the transition at lower energies is highly dependent on the
  calculation of $S$ as illustrated with the alternative energy
  dependence shown in the bottom panel. }
\label{fig3}       
\end{figure}

In Fig.~\ref{fig3} I plot the ridge amplitude as predicted by the
model in Ref.~\cite{DGMV} as dashed curves. For this prediction, the
following parameterization of $\beta$ is used: 
\begin{eqnarray*}
\langle\beta\rangle = (0.655 &+& 0.314\log(N_{binary}/N_{part}))\\  
&\times& (0.05727\log(\sqrt{s_{_{NN}}}) + 0.2933).
\end{eqnarray*}
For the solid curves, I've included an additional ad-hoc transition to
illustrate how the ridge centrality and energy dependence would look
if a sudden decrease in the flow occurs when the energy density in the
collision drops below a critical energy density. This sudden decrease
could easily be masked in data by effects from the hadronic stage. For
this transition, I follow the observation from STAR that the onset of
the ridge occurs at a fixed value of $\frac{1}{S}\frac{dN}{dy}$. I
then extrapolate the observed cutoff to lower energies. This
transition is strongly dependent on the centrality dependence of the
overlap area $S$. For this estimate I use $S=R^2(\theta -
\sin(\theta))$ where $\theta = 2\cos^{-1}(b/2R)$ and $b$ and $R$ are
the impact parameter and nuclear radius repectively. In this case, the
ridge persists only down to $\sqrt{s_{_{NN}}}$ of approximately 39
GeV. Below that, the transverse density is below the minimum even for
the most central collisions. Alternative calculations of $S$ such as
$S=\pi\sqrt{\langle x\rangle^2\langle y\rangle^2}$ yield quite
different results with the ridge persisting to much lower energies.

The flow velocity is only weakly dependent on energy so the variation
of the ridge amplitude with energy is dominated by the change in
$Q_s^2 \propto (\sqrt{s_{_{NN}}})^{0.3}$. This gives a verifiable
prediction for the energy dependence which is independent of the
details of the centrality dependence. This energy dependence can be
checked at LHC and in an upcoming beam energy scan at RHIC. In the
case that the imaging of the flux tubes is made possible by
space-momentum correlations induced by liberated color charges, then
an energy scan of the ridge can be used to map out the phase boundary
of the quark-gluon plasma. Given the observation of an abrupt onset of
the ridge, this project should be a high priority for RHIC.



\section{Beam Energy Scan}
\label{sec:3}

\begin{figure}
\resizebox{0.49\textwidth}{!}{\includegraphics{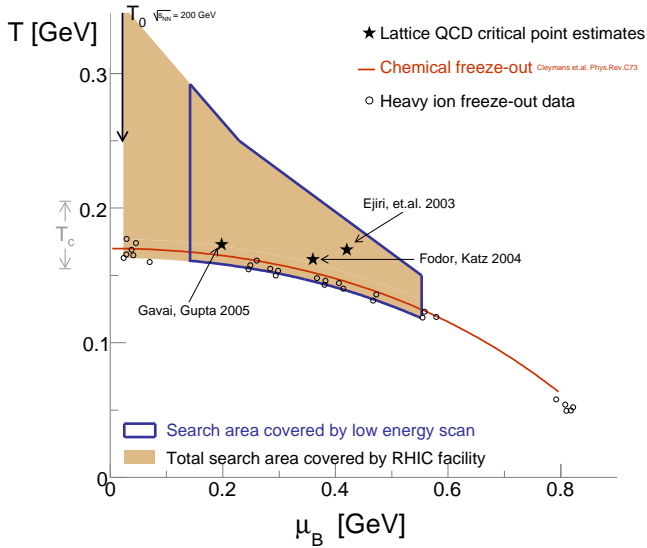}}
\caption{A sketch of the phase diagram of nuclear matter. An estimate
  of the region covered by RHIC is indicated by the tan region. Three
  Lattice estimates for the location of a critical point in the
  diagram are also indicated. }
\label{fig4}       
\end{figure}

A beam energy scan has already been proposed at RHIC. One of the
motivations for the energy scan is to search for signatures of the QCD
critical point. According to lattice QCD, at zero baryon chemical
potential $\mu_B=0$, the transition from hadronic matter below $T_C$
to QGP above $T_C$ is a smooth cross-over. At higher $\mu_{B}$, model
calculations indicate that the transition is a first order phase
transition. The critical point lies where the smooth cross-over
changes to a first order phase transition. See this list of
Refs.~\cite{criticalpoint}. Detecting the presence of the critical
point depends on the ability of experiments to create matter above
$T_C$ at larger and larger $\mu_{B}$. Accessing larger $\mu_{B}$ can
be achieved by decreasing the beam energy. The question of whether the
matter created at high $\mu_{B}$ (low $\sqrt{s_{_{NN}}}$) still
reaches a temperature above $T_C$ is not known. Fig.~\ref{fig4} shows
a sketch of the QCD phase diagram with the region accessible in a RHIC
energy scan outlined in blue. The region covered by RHIC encompasses
several recent estimates for the location of the critical point from
Lattice.

The observation of an abrupt onset for the ridge provides an added
motivation for conducting this energy scan. Several of the promising
explanations for the formation of the ridge indicate that the onset of
the ridge should coincide with the development of large pressure
gradients. This sudden increase in pressure at a particular energy
density could be related to the QCD equation of state and therefore
may be the most direct observation of the EOS accessible in heavy-ion
collisions. The correct physical explanation for the ridge is still
under debate --- but having observed a long range correlation in
heavy-ion collisions which shows an abrupt onset at a given energy
density, the logical next step is to perform an energy scan. In this
scan, the amplitude, longitudinal width, and azimuthal width of the
ridge and critical density for ridge formation can be studied.

\textbf{Acknowledgments:} I'd like to thank Raju Venugopalan and
Adrian Dumitru for help with producing Fig.~\ref{fig3}. I'd like to
thank \'Agnes M\'ocsy, Larry McLerran, Ron Longacre, and Sean Gavin
for many helpful discusions.

%

\end{document}